\documentclass[10pt,final,conference]{IEEEtran}
\IEEEoverridecommandlockouts
\usepackage{cite}
\usepackage{bm}
\usepackage{multicol}
\usepackage{pgfplots}
\usetikzlibrary{shadows, decorations.pathmorphing, shapes.geometric, arrows.meta}
\usepackage{subcaption}
\pgfplotsset{compat=newest}
\pgfplotsset{plot coordinates/math parser=false}
\usepackage{amsmath,amssymb,amsfonts}
\usepackage{algorithmic}
\usepackage{graphicx}
\usepackage{textcomp}
\usepackage{xcolor}
\usepackage{upgreek}
\usepackage{algorithm}
\newcommand{\vect}[1]{\boldsymbol{\mathbf{#1}}}
\newtheorem{assumption}{Assumption}
\DeclareMathOperator{\Tr}{Tr}
\def\BibTeX{{\rm B\kern-.05em{\sc i\kern-.025em b}\kern-.08em
    T\kern-.1667em\lower.7ex\hbox{E}\kern-.125emX}}
\begin{document}

\title{Towards Smarter Sensing: 2D Clutter Mitigation in RL-Driven Cognitive MIMO Radar\\

\thanks{This work was supported in part by the German Federal Ministry of Education and Research (BMBF) in the course of the 6GEM Research Hub under grant 16KISK03 and in part by 6G-ANNA under grant 16KISK095.}
}

\author{
    \IEEEauthorblockN{Adam Umra, Aya Mostafa Ahmed, Aydin Sezgin}
    \IEEEauthorblockA{Institute of Digital Communication Systems, Ruhr University Bochum, Germany
    \\Emails: \{adam.umra, aya.mostafaibrahimahmad, aydin.sezgin\}@rub.de}
}

\maketitle

\begin{abstract}
Motivated by the growing interest in integrated sensing and communication for 6th generation (6G) networks, this paper presents a cognitive Multiple-Input Multiple-Output (MIMO) radar system enhanced by reinforcement learning (RL) for robust multitarget detection in dynamic environments.  The system employs a planar array configuration and adapts its transmitted waveforms and beamforming patterns to optimize detection performance in the presence of unknown two-dimensional (2D) disturbances. A robust Wald-type detector is integrated with a SARSA-based RL algorithm, enabling the radar to learn and adapt to complex clutter environments modeled by a 2D autoregressive process. Simulation results demonstrate significant improvements in detection probability compared to omnidirectional methods, particularly for low Signal-to-Noise Ratio (SNR) targets masked by clutter.
\end{abstract}


\section{Introduction}
Sensing and radar systems are receiving increased research focus, especially with the integration of sensing and communication in 6th generation (6G) networks. Radar enables systems to make informed decisions, improving performance. Cognitive radar (CR) further enhances this by actively sensing and adapting to the environment through a feedback loop between transmitter and receiver modules\cite{sevgi,Hay}. CR systems continuously learn and adapt to their environment using a perception-action cycle, optimizing target detection and tracking in dynamic environments by responding to noise, clutter, and moving targets. While Bayesian filtering is often used in this feedback process, it relies on extensive prior knowledge \cite{Hay}. Reinforcement learning (RL) offers an alternative by learning directly from observations, adaptively selecting optimal actions based on a reward function \cite{sutton}. Multiple-Input Multiple-Output (MIMO) radar systems further enhance CR performance by providing waveform diversity.

MIMO radar systems, unlike traditional phased array systems, offer a higher degrees of freedom (DoF) by transmitting multiple possibly correlated signals. This capability enhances detection by exploiting spatial diversity. MIMO systems are broadly categorized into widely separated~\cite{Haim} and colocated configurations. Colocated MIMO radars, for instance, use closely spaced antennas to achieve significant coherent gains. This is particularly advantageous in CR applications, where real-time beamforming adjustments are critical~\cite{li}.

Previous research on multitarget detection in massive MIMO radar systems using RL~\cite{aya1},~\cite{aya2} has been constrained to one-dimensional antenna arrays and disturbances, limiting their applicability to real-world scenarios. This paper generalizes the framework to a planar MIMO array configuration, designed to operate under unknown two-dimensional (2D) disturbance conditions. Planar arrays are crucial for radars as they offer 2D beam steering, enabling precise target localization in 3D space and enhanced performance in cluttered environments. However, effectively mitigating 2D clutter presents a significant challenge for traditional statistical methods due to the difficulties in accurately modeling its complex and dynamic nature. To address this, a robust Wald-type detector is integrated with a SARSA-based RL algorithm. This approach leverages the learning and adaptation capabilities of RL to dynamically adjust the transmitted waveforms and maximize detection probabilities across range-angle cells, effectively addressing the challenges posed by complex environmental conditions.

\section{SYSTEM MODEL}
We consider a collocated MIMO radar system with $N_T$ transmit antennas and $N_R$ receive antennas. $N=N_T N_R$ denotes the number of spatial channels. Both the transmit and receive arrays are modeled as uniform rectangular planar arrays (UPAs).
\subsection{Transmitted Waveform}
The transmitted signal is modeled as
\begin{equation}
    \vect{x}(t) = \vect{W} \vect{\Psi}(t) \in \mathbb{C}^{N_T},
\end{equation}
where $\vect{\Psi}(t) \in \mathbb{C}^{N_T}$ represents a set of orthonormal basis functions, and $\vect{W} = [\vect{w}_1, \dots, \vect{w}_{N_T}]^T \in \mathbb{C}^{N_T \times N_T}$ is the beamforming weight matrix. The signal $\vect{x}(t)$ is a weighted linear combination of these orthonormal signals, determined by $\vect{W}$. The beamforming matrix $\vect{W}$ satisfies the total transmit power constraint $\Tr\{\vect{W} \vect{W}^H\}= P_T$, where $\Tr\{\cdot\}$ denotes the trace, $\vect{W}^H$ is the Hermitian transpose of $\vect{W}$, and $P_T$ is the total available transmit power.
\subsection{Received Signal}
The complex baseband representation of the received signal, reflected from a point target, is expressed as~\cite{li}
\begin{equation} \label{eq:rs}
    \tilde{\vect{y}}(t) = \alpha \vect{a}_R(\theta, \phi) \vect{a}_T^T(\theta, \phi) \vect{x}(t - \tau) + \tilde{\vect{c}}(t),
\end{equation}
where $\tilde{\vect{y}}(t) \in \mathbb{C}^{N_R}$ is the received signal vector, and $\tau$ is the propagation delay related to the target's range. The transmit and receive array responses are characterized by the steering vectors $\vect{a}_T(\theta, \phi)$ and $\vect{a}_R(\theta, \phi)$, which depend on the target's azimuth angle $\theta$ and elevation angle $\phi$. These steering vectors are generally given by $\vect{a}(\theta, \phi) = \vect{a}_x(\theta, \phi) \otimes \vect{a}_y(\theta, \phi)$, where $\vect{a}_x(\theta, \phi)$ and $\vect{a}_y(\theta, \phi)$ represent the steering vectors along the x- and y-axis, respectively~\cite{Tan}.
The parameter $\alpha \in \mathbb{C}$ is an unknown deterministic scalar that encapsulates the radar cross-section (RCS) and two-way path loss, modeled using the Swerling 0 model~\cite{Swerling}. The vector $\tilde{\vect{c}}(t) \in \mathbb{C}^{N_R}$ models random disturbance, including clutter and additive white Gaussian noise.

At the receiver, the signal is processed with a matched filter $\vect{\Psi}_{\text{MF}}(t)$, synchronized to the estimated delay $\hat{\tau}$. The output of the matched filter is
\begin{equation}
    \vect{Y}(\hat{\tau}) = \int_{0}^{T} \tilde{\vect{y}}(t) \vect{\Psi}^H_{\text{MF}}(t - \hat{\tau}) \, \mathrm{d}t.
\end{equation}
Substituting $\tilde{\vect{y}}(t)$ from Equation \eqref{eq:rs}, we get
\begin{equation} \label{eq:ms}
    \vect{Y}(\hat{\tau}) = \alpha \vect{a}_R(\theta, \phi) \vect{a}_T^T(\theta, \phi) \vect{W} \int_{0}^{T} \vect{\Psi}(t - \tau) \vect{\Psi}^H_{\text{MF}}(t - \hat{\tau}) \, \mathrm{d}t + \vect{C},
\end{equation}
where $\vect{Y} \in \mathbb{C}^{N_R \times N_T}$ is the matched filter output, and $\vect{C} = \int_{0}^{T} \tilde{\vect{c}}(t) \vect{\Psi}^H_{\text{MF}}(t - \hat{\tau}) \, \mathrm{d}t$. Assuming perfect synchronization of the matched filter, i.e., $\hat{\tau} = \tau$, we have
\[
\int_{0}^{T} \vect{\Psi}(t - \tau) \vect{\Psi}^H_{\text{MF}}(t - \hat{\tau}) \, \mathrm{d}t = \vect{I},
\]
where $\mathbf{I}$ is the identity matrix.
Thus, the matched filter output simplifies to
\begin{equation}
    \vect{Y}(\tau) = \alpha \vect{a}_R(\theta, \phi) \vect{a}_T^T(\theta, \phi) \vect{W} + \vect{C}.
\end{equation}

For simplicity, we vectorize the equation as
\begin{equation}
    \vect{y} = \text{vec}(\vect{Y}) = \alpha \vect{h}(\theta, \phi) + \vect{c},
\end{equation}
where $\text{vec}(\cdot)$ denotes the vectorization operator, and $\vect{c} = \text{vec}(\vect{C})$ is the vectorized disturbance. Using the properties of the Kronecker product, the vector $\vect{h}(\theta, \phi)$ is given by
\begin{equation}
    \vect{h}(\theta, \phi) = (\vect{W}^T \vect{a}_T(\theta, \phi)) \otimes \vect{a}_R(\theta, \phi).
\end{equation}
To address the detection problem, it is essential to interpret the received signals over multiple transmissions. Specifically, let the system transmit $K$ pulses, indexed by $k \in \{1, \dots, K\}$, where each pulse $k$ is represented as
\begin{equation}
    \vect{y}^k = \alpha^k \vect{h}^k(\theta, \phi) + \vect{c}^k.
\end{equation}

Finally, we assume accurate sampling along the fast-time (range) dimension, adhering to the Nyquist sampling criterion to avoid aliasing and ensure faithful signal representation~\cite{fried}. This ensures that the matched filtering process correctly captures the target information for reliable range estimation and signal processing.
\subsection{2D Disturbance Model}
Characterizing disturbances in radar systems is challenging due to their often unknown and dynamic nature~\cite{fort2}. To mitigate the risk of model mismatch, we adopt minimal statistical assumptions:
\vspace{1mm} 
\begin{assumption}\label{assump:1}
The disturbance is a stationary, discrete-time, circularly symmetric complex-valued random process with unknown distribution and polynomially decaying autocorrelation.
\end{assumption}
\vspace{1mm} 

This encompasses various practical models, including AR, ARMA, and non-Gaussian correlated processes~\cite{fort1}. The target detection problem is then formulated as a binary hypothesis test.
\subsection{Detection Problem}
The received radar signal is processed by a bank of spatial filters, each targeting a specific angular range. These filters discretize the radar's field of view into angle bins, creating a 2D grid of size $L \times I$, where each bin corresponds to a unique angle pair $(\theta_l, \phi_i)$ for $l = 1, \, \hdots, \, L$ and $i = 1, \, \hdots, \, I$. This 2D approach enables separate tracking of both azimuthal and elevation angles, enhancing spatial resolution compared to previous 1D methods~\cite{aya2}, which typically only tracked a single angular dimension. The signal at each bin $(l, i)$ and each pulse $k$ is represented as
\begin{equation}
\vect{y}^k_{(l,i)} = \alpha^k_{(l,i)} \vect{h}^k_{(l,i)} + \vect{c}^k_{(l,i)}.
\end{equation}

The detection problem in each angle bin is formulated as the following hypothesis test:
\begin{equation}
    \begin{aligned}
        &\mathcal{H}_0: \quad \vect{y}^k_{(l,i)} = \vect{c}^k_{(l,i)}, \quad k=1,\,\hdots,\, K, \\
        &\mathcal{H}_1: \quad \vect{y}^k_{(l,i)} = \alpha^k_{(l,i)} \vect{h}^k_{(l,i)} + \vect{c}^k_{(l,i)}, \quad k=1,\,\hdots,\, K.
    \end{aligned}
\end{equation}
The null hypothesis $\mathcal{H}_0$ assumes that the test cell contains only disturbance, while the alternative hypothesis $\mathcal{H}_1$ represents the presence of a target. The disturbance vector $\vect{c}^k_{(l,i)}$ is modeled as a complex random process with unknown covariance matrix $\vect{\Gamma} = \mathbb{E} \{ \vect{c}^k_{(l,i)} (\vect{c}^k_{(l,i)})^H \}$, satisfying general Assumption \ref{assump:1}. The detection is performed on a per-pulse basis, as the number of targets, corresponding angles, and signal-to-noise ratio (SNR) can vary with each pulse. Additionally, the disturbance statistics may change over time and space, so we assume a single snapshot scenario for analysis.

To distinguish between $\mathcal{H}_0$ and $\mathcal{H}_1$, a test statistic $\Lambda\left(\vect{y}^k_{(l,i)}\right)$ is employed, where the decision rule is defined as
\begin{equation}\label{eq:hyp}
    \Lambda\left(\vect{y}^k_{(l,i)}\right) \underset{\mathcal{H}_0}{\overset{\mathcal{H}_1}{\gtrless}} \delta.
\end{equation}
In radar applications, controlling the false alarm probability $P_{\text{FA}}$ is crucial. Therefore, the threshold $\delta$ must be set to satisfy
\begin{equation}\label{eq:PFA}
    \text{Pr}\left(\Lambda(\vect{y}^k_{(l,i)}) > \delta \, |\, \mathcal{H}_0\right) =  \int\displaylimits_{\delta}^{\infty} f_{\Lambda  | \mathcal{H}_0}(a \,|\, \mathcal{H}_0) \,\mathrm{d}a = P_{\text{FA}},
\end{equation}
where $f_{\Lambda  | \mathcal{H}_0}$ is the probability density function (pdf) of the test statistic $\Lambda(\vect{y}^k_{(l,i)})$ under the null hypothesis $\mathcal{H}_0$.

In typical radar systems, traditional model-based detectors like the Generalized Likelihood Ratio Test (GLRT) or the Wald test are used to solve (\ref{eq:hyp}). However, these methods are not directly applicable in this scenario due to the unknown functional form of the pdf of $\vect{c}^k_{(l,i)}$. To address this, a robust Wald-type detector is applied~\cite{fort1}, which only requires the disturbance model to satisfy Assumption \ref{assump:1}.

The test statistic for the robust Wald detector is given by
\begin{equation}\label{eq:detector}
    \Lambda^k_{{(l,i)},\text{RW}} = \frac{2 \left|(\vect{h}^k_{(l,i)})^H \vect{y}^k_{(l,i)} \right|^2}{(\vect{h}^k_{(l,i)})^H \hat{\vect{\Gamma}} \vect{h}^k_{(l,i)}},
\end{equation}
where $\hat{\vect{\Gamma}}$ is the estimated covariance matrix of the disturbance. Further details on the computation of $\hat{\vect{\Gamma}}$ and the asymptotic distribution of $\Lambda^k_{{(l,i)},\text{RW}}$ can be found in~\cite{fort1}. The threshold $\delta$ in (\ref{eq:hyp}), which guarantees the desired $P_{\text{FA}}$ irrespective of the disturbance distribution, is given by $\delta = H^{-1}_{\chi^2_2}(1-P_{\text{FA}})$, where $H^{-1}_{\chi^2_2}(\cdot)$ is the inverse cumulative distribution function (cdf) of a chi-squared distribution with 2 degrees of freedom, defined as
\begin{equation}
    H_{\chi^2_2}(\delta) = \int_{-\infty}^{\delta} p_{\Lambda^k_{{(l,i)},\text{RW}}}(a|\mathcal{H}_0) \,\mathrm{d}a.
\end{equation}

Having established the detection problem and a robust detection strategy, we now explore how RL can be integrated to further enhance the radar's performance in dynamic and uncertain environments.
\section{RL-Based Cognitive Radar}
Reinforcement Learning (RL) is a subset of machine learning enabling an agent to achieve a task by iteratively interacting with its environment~\cite{sutton}. At each time step \(k\), the agent evaluates an action \(a_k\) based on two key inputs: the current state of the environment, \(s_k\), and the reward, \(r_k\), received after taking that action. The state \(s_k\) represents the agent’s perception of the environment at time \(k\), while \(r_k\) quantifies the immediate benefit or cost of action \(a_k\) in \(s_k\). Through repeated interactions, the agent learns a policy—a mapping from states to actions—aiming to maximize long-term rewards by identifying optimal actions in various states.

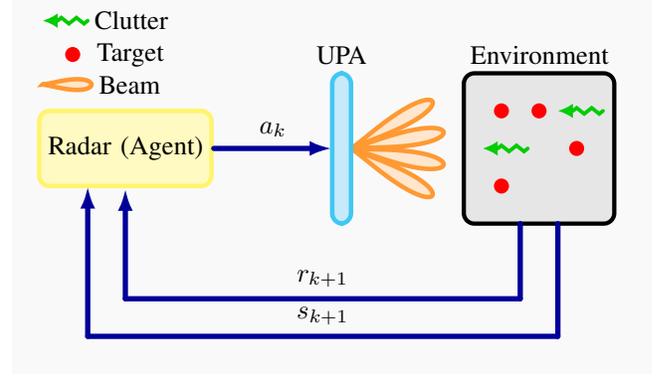
\begin{figure}[htbp]
    \centering
    \begin{tikzpicture}[line cap=round]
    \begin{scope}[scale=0.5]  
        
        \fill[gray!5!white] (-9,-6) rectangle (8,4); 
        
        \foreach \x in {30, 10, -10, -30} {
            \fill[orange!20!white, opacity=0.6] plot[variable=\t,domain=0:360,smooth,samples=50] ({\x+8*sin(\t)}:{2.5*pow(sin(\t/2),3)}) -- cycle;
            \draw[orange!80, line width=0.06cm, smooth] plot[variable=\t,domain=0:360,samples=50] ({\x+8*sin(\t)}:{2.5*pow(sin(\t/2),3)});
        }

        \fill[cyan!20!white, rounded corners] (-0.5,-2) rectangle (0,2);
        \draw[cyan!60, rounded corners, line width=0.06cm] (-0.5,-2) rectangle (0,2);
        \node[above] at (-0.25,2) {UPA}; 
    
        \draw[ultra thick, fill=gray!20!white, rounded corners] (3,-2) rectangle (7,2);
        \node[above] at (5,2) {Environment}; 

        \foreach \x/\y in {5/1, 4/-1, 6/0, 4/1} {
            \fill[red, opacity=0.9] (\x,\y) circle (0.2cm); 
        }

        \draw[line width=0.05cm, -{Latex[length=2mm, width=2mm]}, draw=green!80!black, decorate, decoration={zigzag,amplitude=.4mm,segment length=2mm,post length=2mm}] (6.7,1) -- (5.5,1); 
        \draw[line width=0.05cm, -{Latex[length=2mm, width=2mm]}, draw=green!80!black, decorate, decoration={zigzag,amplitude=.4mm,segment length=2mm,post length=2mm}] (4.7,0) -- (3.5,0); 

        \draw[line width=0.05cm, -{Latex[length=2mm, width=2mm]}, draw=green!80!black, decorate, decoration={zigzag,amplitude=.4mm,segment length=2mm,post length=2mm}] (-7,3.4) -- (-8.2,3.4) node[right] at (-7.1,3.4) {Clutter}; 
        \fill[red, opacity=0.9] (-7.4,2.5) circle (0.2cm) node[black,right] at (-7,2.5) {Target}; 
        
        \begin{scope}[shift={(-8.2,1.7)}] 
            \fill[orange!20!white, opacity=0.6] plot[variable=\t, domain=0:360, smooth, samples=50] ({11*sin(\t)}:{1.3*pow(sin(\t/2),3)}) --cycle;
            \draw[orange!80, line width=0.06cm, smooth] plot[variable=\t, domain=0:360, samples=50] ({11*sin(\t)}:{1.3*pow(sin(\t/2),3)}) node[black,right] at (1.2,0) {Beam};;
        \end{scope}

        \node[rectangle, draw=yellow!70, fill=yellow!30!white, minimum height=1cm, minimum width=2cm, line width=0.06cm, rounded corners] (r) at (-6,0) {Radar (Agent)};

        \draw[line width=0.07cm, -{Latex[length=3mm, width=2mm]}, draw=blue!60!black] (r) -- (-0.5,0) node[midway, above] {$a_k$};
        \draw[line width=0.07cm, draw=blue!60!black] (5.5,-2) -- (5.5,-5);
        \draw[line width=0.07cm, draw=blue!60!black] (5.5,-5) -- (-7,-5) node[midway, above] {$s_{k+1}$};
        \draw[line width=0.07cm, -{Latex[length=3mm, width=2mm]}, draw=blue!60!black] (-6,-4) -- (r.south);
        \draw[line width=0.07cm, draw=blue!60!black] (4.5,-2) -- (4.5,-4);
        \draw[line width=0.07cm, draw=blue!60!black] (4.5,-4) -- (-6,-4) node[midway, above] {$r_{k+1}$};
        \draw[line width=0.07cm, -{Latex[length=3mm, width=2mm]}, draw=blue!60!black] (-7,-5) -- (-7,-1);

    \end{scope} 
\end{tikzpicture}
    \caption{RL-based MIMO radar system with UPA forming beams for target tracking. The radar agent sends actions to the UPA, receiving feedback in the form of state and reward signals for adaptive tracking via reinforcement learning.}
    \label{fig:sys}
\end{figure}

In our detection framework, the agent is represented by the MIMO radar system, tasked with detecting multiple targets in an environment with unknown disturbances, as illustrated in Fig. \ref{fig:sys}. The radar (as the RL agent) iteratively improves its detection strategy by adjusting actions based on observed states and rewards, thereby enhancing its detection capabilities in unpredictable conditions~\cite{aya2, fort1}.

\subsection{State Definition}
The environment's current status is defined by state \( s_k \), constructed from the statistic \( \Lambda^k_{{(l,i)},\text{RW}} \), with
\begin{equation} \label{eq:thres}
    \bar{\Lambda}^k_{{(l,i)}}  = 
    \begin{cases}
        1 & \Lambda^k_{{(l,i)},\text{RW}} > \delta,\\
        0 & \text{otherwise},
    \end{cases}
\end{equation}
where $\bar{\Lambda}^k_{{(l,i)}}$ indicates whether an angle bin $(l,i)$ at time $k$ contains a target. Therefore, $s_k$ represents the count of target-containing angle bins at $k$
\begin{equation}\label{eq:states}
    s_k = \sum_{l=1}^{L}\sum_{i=1}^{I} \bar{\Lambda}^k_{{(l,i)}}.
\end{equation}
The set of possible states is then \( S = \{0, \dots, M\} \), where \( M \) is the maximum number of detectable targets.

\subsection{Action Selection}
As the RL agent, the MIMO radar selects action \( a_k \) based on \( s_k \). Each action comprises two parts: 1) selecting angle bins likely to contain targets based on observed state, and 2) optimizing the beamforming matrix \( \vect{W} \) to concentrate power toward these bins, maximizing detection likelihood. 

Actions are chosen from $\mathcal{A} = \{\Theta_m \mid m \in \{0, 1, \dots, M\}\}$, where each $\Theta_m$ represents $m$ candidate angle bins $(\hat{\theta}_1, \hat{\phi}_1), \dots, (\hat{\theta}_m, \hat{\phi}_m)$, based on the highest $m$ values of $\Lambda^k_{{(l,i)},\text{RW}}$ in (\ref{eq:detector}). Note that the indices do not imply that the angles are identical; rather, they indicate that these angle bins correspond to different targets. Given $\Theta_m$, $\vect{W}$ is optimized to enhance transmit power toward the selected bins by solving
\begin{equation}\label{eq:opt}
    \begin{aligned}
        \max_{\vect{W}} \min_{j \in \mathcal{T}_m} \quad & \vect{a}_T^T(\hat{\theta}_j,\hat{\phi}_j) \vect{W} \vect{W}^H \vect{a}_T^*(\hat{\theta}_j,\hat{\phi}_j)\\
        \text{s.t.} \quad & \Tr(\vect{W} \vect{W}^H) = P_T,
    \end{aligned}
\end{equation}
where \( \mathcal{T}_m = \{1, \dots, m\} \) and \( (\hat{\theta}_j, \hat{\phi}_j) \in \Theta_m \). Using maximum power design~\cite{stoica}, the solution for \( \vect{W} \) is
\begin{equation}
    \vect{W} = P_T^{1/2} \frac{\vect{v}_{\max}}{\|\vect{v}_{\max}\|}
\end{equation}
where \( \vect{v}_{\max} \) is the eigenvector of the largest eigenvalue of
\begin{equation}
    \vect{A} = \sum_{j \in \mathcal{T}_m} \vect{a}_T(\hat{\theta}_j, \hat{\phi}_j) \vect{a}_T^H(\hat{\theta}_j, \hat{\phi}_j),
\end{equation}
ensuring \( \Tr(\vect{W} \vect{W}^H) = P_T \).

\subsection{Reward Calculation}
The reward $\hat{P}_{D,{(l,i)}}^k$, representing detection performance feedback at each \( k \), is calculated asymptotically as $N \rightarrow \infty$~\cite{fort1}
\begin{equation}
    \hat{P}_{D,{(l,i)}}^k = Q_1 \left(\sqrt{\hat{\zeta}^k_{(l,i)}}, \sqrt{\delta}\right),
\end{equation}
where  $Q_1 (\cdot,\cdot)$ is the first order Marcum Q function~\cite{nuttall},
\begin{equation}
    \hat{\zeta}^k_{({l,i})} = \frac{2|\hat{\alpha}^k_{({l,i})}|^2 \| \vect{h}^k_{({l,i})} \|^4}{(\vect{h}^k_{({l,i})})^H \vect{\hat{\Gamma}}_{({l,i})} \vect{h}^k_{({l,i})}},
\end{equation}
and
\begin{equation}
    \hat{\alpha} = \frac{(\vect{h}^k_{({l,i})})^H  \vect{y}^k_{({l,i})}}{\| \vect{h}^k_{({l,i})} \|}.
\end{equation}
The cumulative reward at \( k \) is given as
\begin{equation}\label{eq:reward}
    r_{k+1} = \sum_{(l,i) \in \mathcal{B}_\text{target}} \hat{P}_{D,{(l,i)}}^k - \sum_{(l,i) \in \mathcal{B}_\text{non-target}} \hat{P}_{D,{(l,i)}}^k,
\end{equation}
where \( \mathcal{B}_\text{target} \) and \( \mathcal{B}_\text{non-target} \) indicate bins with and without likely targets, respectively.

\subsection{SARSA Algorithm for Detection}
SARSA, which stands for state-action-reward-state-action, is an algorithm in reinforcement learning that updates the Q-function at each $k$ based on the sequence $s_k, a_k, r_{k+1}, s_{k+1}, a_{k+1}$~\cite{Poole}. 
\begin{algorithm}[htbp]
\caption{SARSA for MIMO Target Detection}
\textbf{Initialize} $\vect{Q} = 0_M$, state $s_0 = 1$, action $a_0 = 1$, $K = 50$, and $\vect{W}_k = I$

\textbf{repeat} for each time step $k$:
\begin{algorithmic}[1]
    \STATE Take action $a_k$ by transmitting waveform (\ref{eq:rs}) using $\vect{W}_k$
    \STATE Acquire received signal $\vect{y}_{(l,i)}^k, \forall l, i$
    \STATE Calculate $s_{k+1}$ from (\ref{eq:states}) and the reward $r_{k+1}$ as in (\ref{eq:reward})
    \STATE Choose action $a_{k+1}$ using $\epsilon$ greedy, identify $\theta_m$ and $\mathcal{T}_m$
    \STATE Update $Q(s_k, a_k)$ using the update rule in (\ref{eq:qfunc})
    \STATE $s_k \leftarrow s_{k+1}; \quad a_k \leftarrow a_{k+1}$
    \IF{$s_{k+1} \neq 0$}
        \STATE Solve for $\vect{W}_{k+1}$ in (\ref{eq:opt})
    \ELSE
        \STATE $\vect{W}_k = I$
    \ENDIF
\end{algorithmic}
\textbf{until} Observation time ends
\label{alg:alg1}
\end{algorithm}
The Q-function approximates cumulative reward starting from state $s_k$ with action $a_k$ under policy $\pi$. The agent updates a matrix $\vect{Q} \in \mathbb{R}^{(M+1) \times (M+1)}$ with elements given by
\begin{align}\label{eq:qfunc}
    Q(s_k, a_k) \leftarrow \; & Q(s_k, a_k) \;+ \\
    &\alpha \left( r_{k+1} + \gamma \,Q(s_{k+1}, a_{k+1}) - Q(s_k, a_k) \right),  \nonumber
\end{align}
where $\alpha$ and $\gamma$ are the learning rate and discount factor, respectively.
The radar uses an $\epsilon$-greedy policy~\cite{sutton} to determine $a_k$ by defining the size of $\Theta_m$ (i.e $m$). With a probability of $1-\epsilon$, it chooses the optimal action $ a_{\text{opt}} \triangleq \arg \max_{a \in \mathcal{A}} Q(s_{k+1}, a)$. However, sometimes it takes a chance and chooses a random action with probability $\epsilon$ to explore new possibilities. This balance helps the radar learn and adapt effectively. A detailed SARSA operation is shown in Algorithm \ref{alg:alg1}.  
\section{SIMULATION RESULTS}
The simulations consider $L=I=20$ angle bins with a total of $20\times 20$  angle bins. The inter-element spacing is set as $d_x = d_y = \frac{\lambda}{2}$. Thus, the spatial frequency is $\nu_x = \frac{1}{2}\sin(\theta)\cos(\phi)$ and $\nu_y = \frac{1}{2}\sin(\theta)\sin(\phi)$ and the steering vectors result in $\vect{a}_x = [1,e^{j2\pi\nu_x},\dots,e^{j2\pi(\sqrt{N_T}-1)\nu_x}]^T$ and $\vect{a}_y = [1,e^{j2\pi\nu_y},\dots,e^{j2\pi(\sqrt{N_T}-1)\nu_y}]^T$ for transmitter and receiver. The grid is expressed in terms of the spatial frequency $\vect{\nu}_{x} = \vect{\nu}_{y} = [-0.5:0.45]$ The disturbance vector $\mathbf{c}_{l,i}^k$ is modeled as a circular complex AR process~\cite{dudgeon}, where dependencies are introduced along each dimension independently in a separable fashion:
\begin{equation}
    c_{n_{x}, n_{y}} = \sum_{i=1}^{p} \rho_{x, i} c_{n_{x} - i, n_{y}} + \sum_{j=1}^{q} \rho_{y, j}  c_{n_{x}, n_{y} - j} + w_{n_{x}, n_{y}},
\end{equation}
where \( c_{n_{x}, n_{y}} \) denotes the disturbance at the spatial index \( (n_{x}, n_{y}) \) in 2D space, \( \rho_{x, i} \) and \( \rho_{y, j} \) are the AR coefficients along each dimension, and \( w_{n_{x}, n_{y}} \) represents the driving noise at the spatial location \( (n_{x}, n_{y}) \). By introducing dependencies independently along each spatial dimension, this process utilizes \( p+q \) coefficients rather than the \( p \times q \) coefficients required for full 2D dependency modeling. Consequently, dependencies across the grid are captured without requiring all possible 2D combinations.

Similar to the 1D model in~\cite{aya2}, the driving noise \( w_{n_{x}, n_{y}} \) follows a heavy-tailed t-distribution to account for non-Gaussian characteristics commonly observed in impulsive or cluttered radar environments:
\begin{equation}
    p_{w}(w_{n_x, n_y}) = \frac{\mu}{\sigma^2_w} \left( \frac{\mu}{\xi} \right)^\mu \left( \frac{\mu}{\xi} + \frac{|w_{n_x,n_y}|^2}{\sigma^2_w} \right)^{-(\mu+1)},
\end{equation}
where \( \mu \) determines the tail heaviness and \( \sigma^2_w \) is the variance of the noise. The power spectral density (PSD) in 2D represents the distribution of disturbance power across spatial frequencies:
\begin{equation}
     S(\nu_{x}, \nu_{y}) = \sigma^2_w \left| 1 - \sum_{n=1}^{p}  \rho_{x, n} e^{-j 2 \pi \nu_{x} n} \sum_{n=1}^{q} \rho_{y, n} e^{-j 2 \pi \nu_{y} n}\right|^{-2}.
\end{equation}

\begin{figure}
    \centering
    \input{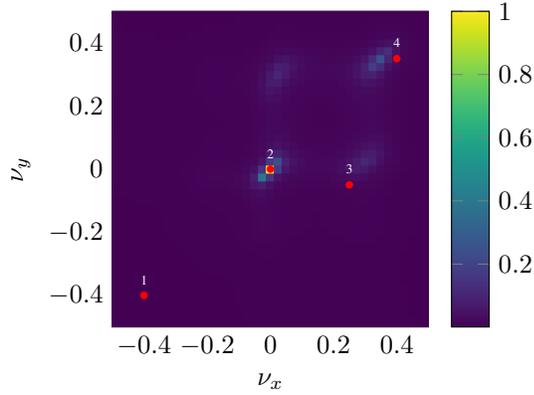}
    \caption{Disturbance PSD along with targets locations as red circles.}
    \label{fig:psd}
\end{figure}

 We compare the results of our RL-based waveform selection method with a standard approach where the transmission is omnidirectional and power is distributed evenly. In this baseline method, orthonormal waveforms are sent, with the total transmission power equally shared among the antennas, adhering to a total power limit of \(P_T = 1\). We are considering five targets located at $\nu = \{(-0.4,-0.4), (0,0), (0.25,-0.05), (0.4,0.35)\}$ with $\text{SNR} =[-5\text{dB},-8\text{dB}, -10\text{dB}, -9\text{dB}]$. The innovation process prameters are chosen to be $\lambda = 2$ and $\sigma_w^2 = 1$. 
The AR disturbance is modeled with the order $p = q = 6$, with the coefficient vectors 
\begin{align}
    \rho_{x} =\rho_{y} = & [0.5e^{-j2\pi0.4},0.6e^{-j2\pi0.2},0.7e^{-j2\pi0}, \nonumber\\   
    & 0.4e^{-j2\pi0.1},0.5e^{-j2\pi0.3},0.6e^{-j2\pi0.35}]^T.
\end{align}
The disturbance PSD is shown in Fig. \ref{fig:psd}, where the targets are marked as red circles. Note that the disturbance PSD has multiple peaks. The next sections discuss how the algorithm performs for different aspects. 

\subsection{Stationary Environment}
\begin{figure}[htbp]
    \centering
    \begin{subfigure}[b]{1\linewidth}
        \centering
\begin{tikzpicture}
    \begin{axis}[
        x dir=reverse,
        y dir=reverse,
        view={45}{25},
        width=5.8cm,
        height=5.8cm,
        z buffer=sort,
        xmin=0,xmax=20,
        ymin=-0,ymax=20,
        zmin=0,zmax=50,
        enlargelimits=upper,
        xtick={3,7,...,19},
        xticklabels={-0.4,-0.2,0,0.2,0.4},
        ytick={3,7,...,19},
        yticklabels={-0.4,-0.2,0,0.2,0.4},
        ztick={0,10,...,50},
        tick label style={font=\small},
        xlabel={$\nu_x$},
        ylabel={$\nu_y$},
        zlabel={$\text{Time}$},
        point meta=\thisrowno{3},
        axis background/.style={fill={rgb,255:red,68; green,1; blue,84}},
        point meta min = 0,
        point meta max = 1,
        colormap name=viridis,
        colorbar,
        ]
        \addplot3 [only marks,scatter,mark=cube*,mark size=7,  cube/size z={2pt},cube/size x={4pt},cube/size y={4pt},scatter/use mapped color=
        {draw=mapped color,fill=mapped color}
        ]
            table {pgfplots_scatterdata4.dat};
    \end{axis}
\end{tikzpicture}%
        \caption{RL-based beamforming}
        \label{fig:RLbased}
    \end{subfigure}
    \par\bigskip
    \begin{subfigure}[b]{1\linewidth}
        \centering
\begin{tikzpicture}
    \begin{axis}[
        x dir=reverse,
        y dir = reverse,
        view={45}{25},
        width=6cm,
        height=6cm,
        z buffer=sort,
        xmin=0,xmax=20,
        ymin=-0,ymax=20,
        zmin=0,zmax=50,
        enlargelimits=upper,
        xtick={3,7,...,19},
        xticklabels={-0.4,-0.2,0,0.2,0.4},
        ytick={3,7,...,19},
        yticklabels={-0.4,-0.2,0,0.2,0.4},
        tick label style={font=\small},
        ztick={0,10,...,50},
        ztick={0,10,...,50},
        xlabel={$\nu_x$},
        ylabel={$\nu_y$},
        zlabel={$\text{Time}$},
        point meta=\thisrowno{3},
        point meta min = 0,
        point meta max = 1,
        axis background/.style={fill={rgb,255:red,68; green,1; blue,84}},
        colormap name=viridis,
        colorbar
        ]
        \addplot3 [only marks,scatter,mark=cube*,mark size=7,  cube/size z={2pt},cube/size x={4pt},cube/size y={4pt}, scatter/use mapped color=
        {draw=mapped color,fill=mapped color}
        ]
            table {pgfplots_scatterdata1.dat}; 
    \end{axis}
\end{tikzpicture}%
        \caption{Omnidirectional with equal power allocation}
        \label{fig:omni}
    \end{subfigure}
    \caption{Detection performance of RL beamforming versus omnidirectional with equal power allocation under $P_{FA} = 10^{-5}$ and
$N = N_TN_R = 10^4$.}
    \label{fig:perf}
\end{figure}
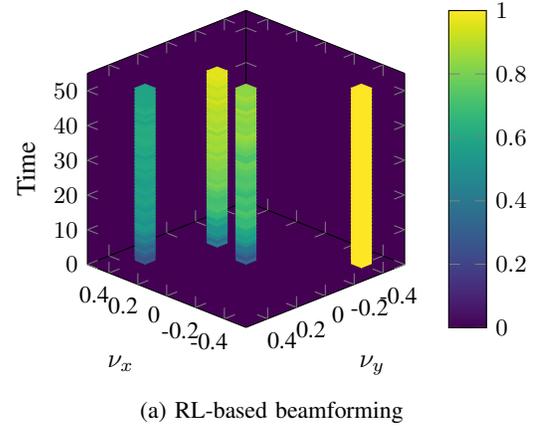
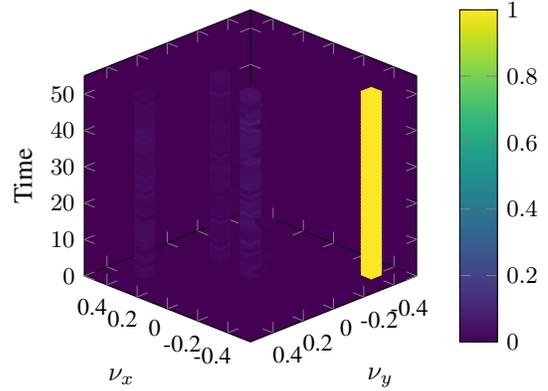
To exploit the benefits of RL, we compare the proposed RL-based beamformer against the omnidirectional approach with no RL. We analyze the performance of the algorithm for a MMIMO regime where $N = N_T N_R = 10^4$ (i.e. $N_T = N_R = 100$) and $P_{FA} = 10^{-5}$. Fig. \ref{fig:perf} shows the detection capability of our algorithm against the omnidirectional approach. For this, we calculated the threshold in (\ref{eq:thres}) within each time step, then the average is taken across 200 Monte Carlo runs. Fig. \ref{fig:RLbased} demonstrates better detection performance for all targets even the ones with low SNR . The algorithm shows that it learns over time. In the first ten time steps, the agent works on understanding the disturbance, gradually building its experience as time goes on. Conversely, in the omnidirectional approach in Fig. \ref{fig:omni}, the targets with lower SNR are mostly masked under the disturbance peaks and can not be detected.

\subsection{Varying Spatial Channels}
\begin{figure}[htbp]
    \centering
    \begin{subfigure}[b]{1\linewidth}
        \centering
        \definecolor{mycolor1}{rgb}{0.00000,0.44700,0.74100}
\definecolor{mycolor2}{rgb}{0.85000,0.32500,0.09800}

\begin{tikzpicture}

\begin{axis}[
width=6cm,
height=1.6cm,
scale only axis,
xmode=log,
xmin=81,
xmax=10000,
xminorticks=true,
xlabel style={font=\color{white!15!black}},
xlabel={Spatial Channels N},
ymin=0,
ymax=1,
ylabel style={font=\color{white!15!black}},
ylabel={$P_D$},
axis background/.style={fill=white},
xmajorgrids,
xminorgrids,
ymajorgrids,
legend style={legend cell align=left, draw=white!15!black, at={(0.6,0.8)},anchor=north west}
],
\addplot [color=mycolor1, mark=o, mark options={solid, mycolor1}, style={ultra thick}]
  table[row sep=crcr]{
81	0.214\\
225	0.14\\
625	0.252\\
1296	0.998\\
2401	0.988\\
4096	1\\
6561	1\\
10000	1\\
};
\addlegendentry{orth}

\addplot [color=mycolor2, mark=square, mark options={solid, mycolor2},style={ultra thick}]
  table[row sep=crcr]{
81	0.524\\
225	0.844\\
625	0.95\\
1296	1\\
2401	1\\
4096	1\\
6561	1\\
10000	1\\
};
\addlegendentry{RL}

\end{axis}
\end{tikzpicture}%
        \caption{$\nu_1 = (-0.4,-0.4)$}
        \label{fig:varN1}
    \end{subfigure}
    \par\bigskip
    \begin{subfigure}[b]{1\linewidth}
        \centering
        \definecolor{mycolor1}{rgb}{0.00000,0.44700,0.74100}
\definecolor{mycolor2}{rgb}{0.85000,0.32500,0.09800}

\begin{tikzpicture}

\begin{axis}[
width=6cm,
height=1.6cm,
scale only axis,
xmode=log,
xmin=81,
xmax=10000,
xminorticks=true,
xlabel style={font=\color{white!15!black}},
xlabel={Spatial Channels N},
ymin=0,
ymax=1,
ylabel style={font=\color{white!15!black}},
ylabel={$P_D$},
axis background/.style={fill=white},
xmajorgrids,
xminorgrids,
ymajorgrids,
legend style={legend cell align=left, draw=white!15!black, at={(0.1,0.9)},anchor=north west}
],
\addplot [color=mycolor1, mark=o, mark options={solid, mycolor1}, style={ultra thick}]
  table[row sep=crcr]{
81	0.01\\
225	0.004\\
625	0.004\\
1296	0.006\\
2401	0.006\\
4096	0.014\\
6561	0.024\\
10000	0.044\\
};
\addlegendentry{orth}
\addplot [color=mycolor2, mark=square, mark options={solid, mycolor2},style={ultra thick}]
  table[row sep=crcr]{
81	0.004\\
225	0.02\\
625	0.132\\
1296	0.26\\
2401	0.454\\
4096	0.702\\
6561	0.736\\
10000	0.804\\
};
\addlegendentry{RL}
\end{axis}
\end{tikzpicture}%
        \caption{$\nu_2 = (0,0)$}
        \label{fig:varN3}
    \end{subfigure}
    \caption{$P_D$ using RL and alternative approaches of existing targets across different virtual antenna array size}
    \label{fig:varN}
\end{figure}
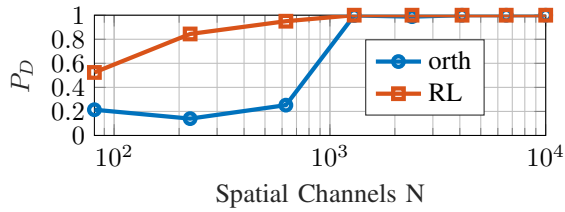
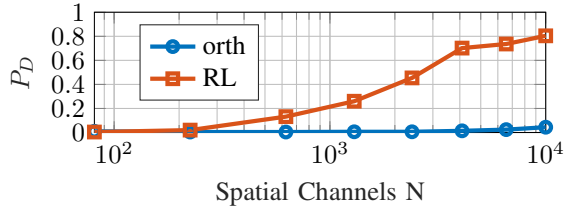

In Fig. \ref{fig:varN}, the detection probability $P_D$ is illustrated as a function of the virtual spatial channels $N$, with a preset false alarm probability $P_{FA} = 10^{-5}$ for targets positioned at $\nu_1 = (-0.4, -0.4)$ in Fig. \ref{fig:varN1} and $\nu_2 = (0, 0)$ in Fig. \ref{fig:varN3}. These target locations were specifically chosen to highlight different detection challenges: target two is masked by a clutter peak, while target one remains unobscured. The number of transmit and receive antennas is given by $N_T = N_R = [9, 16, 25, 36, 49, 64, 81, 100]$, where $N = N_T N_R$. Notably, as $N$ approaches $10^4$ (i.e., $N_T = N_R = 100$), the algorithm achieves successful detection of target two, even under clutter interference, contrasting with the performance of the omnidirectional approach. Additionally, for target one, omnidirectional power allocation reliably detects the target with a large number of antennas, given its high SNR and lack of clutter interference. It is also noteworthy that $P_D$ jumps significantly from 0.2 for $N_T = N_R = 25$ to nearly 1 for $N_T = N_R = 36$ and higher, as the array size crosses a detection threshold where the SNR and clutter suppression become sufficient for reliable detection.
\subsection{Varying SNR}
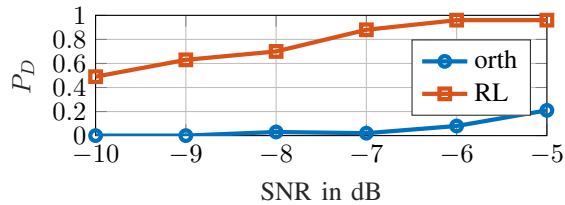
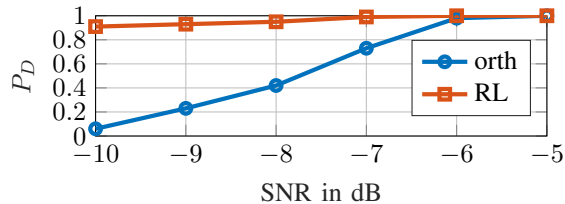
\begin{figure}[htbp]
    \centering
    \begin{subfigure}[b]{1\linewidth}
        \centering
%
%
\definecolor{mycolor1}{rgb}{0.00000,0.44700,0.74100}%
\definecolor{mycolor2}{rgb}{0.85000,0.32500,0.09800}%
\begin{tikzpicture}

\begin{axis}[%
width=6cm,
height=1.6cm,
scale only axis,
xmin=-10,
xmax=-5,
xlabel style={font=\color{white!15!black}},
xlabel={SNR in dB},
ymin=0,
ymax=1,
ylabel style={font=\color{white!15!black}},
ylabel={$P_D$},
axis background/.style={fill=white},
xmajorgrids,
ymajorgrids,
legend style={legend cell align=left, draw=white!15!black, at={(0.7,0.8)},anchor=north west}
]
\addplot [color=mycolor1, mark=o, mark options={solid, mycolor1}, style={ultra thick}]
  table[row sep=crcr]{
-5	0.21\\
-6	0.08\\
-7	0.02\\
-8	0.03\\
-9	0\\
-10	0\\
};
\addlegendentry{orth}

\addplot [color=mycolor2, mark=square, mark options={solid, mycolor2},style={ultra thick}]
  table[row sep=crcr]{
-5	0.96\\
-6	0.96\\
-7	0.88\\
-8	0.7\\
-9	0.63\\
-10	0.49\\
};
\addlegendentry{RL}
\end{axis}
\end{tikzpicture}%
        \caption{$N = 625$ (i.e $N_T = N_R = 25$)}
        \label{fig:snr25}
    \end{subfigure}
    \par\bigskip
    \begin{subfigure}[b]{1\linewidth}
        \centering
%
%
\definecolor{mycolor1}{rgb}{0.00000,0.44700,0.74100}%
\definecolor{mycolor2}{rgb}{0.85000,0.32500,0.09800}%
\begin{tikzpicture}

\begin{axis}[%
width=6cm,
height=1.6cm,
scale only axis,
xmin=-10,
xmax=-5,
xlabel style={font=\color{white!15!black}},
xlabel={SNR in dB},
ymin=0,
ymax=1,
ylabel style={font=\color{white!15!black}},
ylabel={$P_D$},
axis background/.style={fill=white},
xmajorgrids,
ymajorgrids,
legend style={legend cell align=left, draw=white!15!black, at={(0.7,0.8)},anchor=north west}
]
\addplot [color=mycolor1, mark=o, mark options={solid, mycolor1}, style={ultra thick}]
  table[row sep=crcr]{
-5	1\\
-6	0.98\\
-7	0.73\\
-8	0.42\\
-9	0.23\\
-10	0.06\\
};
\addlegendentry{orth}

\addplot [color=mycolor2, mark=square, mark options={solid, mycolor2},style={ultra thick}]
  table[row sep=crcr]{
-5	1\\
-6	1\\
-7	0.99\\
-8	0.95\\
-9	0.93\\
-10	0.91\\
};
\addlegendentry{RL}
\end{axis}
\end{tikzpicture}%
        \caption{$N = 1296$ (i.e $N_T = N_R = 36$)}
        \label{fig:snr36}
    \end{subfigure}
    \caption{$P_D$ using RL and alternative approaches over SNR for target at position $\nu_1 = (-0.4,-0.4)$}
    \label{fig:varSNR}
\end{figure}

The results in Fig. \ref{fig:varSNR} illustrate detection probability $P_D$ across varying SNR for two antenna configurations: $N = 625$ (with $N_T = N_R = 25$) and $N = 1269$ (with $N_T = N_R = 36$). These setups reveal the sensitivity of detection under omnidirectional beamforming to SNR variations and highlight the effectiveness of a larger array size. For the omnidirectional case, $P_D$ improves significantly when $N = 1269$ compared to $N = 625$. Specifically, with $N = 1269$, the system benefits from enhanced target differentiation and clutter suppression, which is especially beneficial for targets with low SNR positioned close to clutter peaks. Additionally, the results indicate a steep improvement in $P_D$ when the array size increases from 625 to 1269 elements. This threshold effect aligns with findings in Fig. \ref{fig:varN}, where the increased number of spatial channels at $N_T = N_R = 36$ allows the system to maintain a higher detection probability despite low SNR conditions, effectively pushing the performance boundary of the omnidirectional case towards that of the RL-based adaptive beamforming approach. This confirms the strong positive impact of larger array sizes on detection probability, particularly under challenging SNR and clutter conditions.
\section{CONCLUSION}
In this paper, we introduced a cognitive MIMO radar system that leverages RL to enhance multitarget detection under complex 2D disturbances using a UPA. The proposed RL-based approach dynamically adapts beamforming and waveform selection, optimizing detection performance despite non-Gaussian and non-stationary clutter. Our simulations demonstrate significant improvements over omnidirectional methods, especially in clutter-dense scenarios with low-SNR targets. Notably, the results confirm that the detection performance achieved in the one-dimensional ULA disturbance model extends effectively to the UPA case, validating the robustness of our approach across different array configurations. This advancement highlights the potential of RL-driven cognitive radar systems to adapt and excel in dynamic environments.



\begin{thebibliography}{99}
\bibitem{sevgi} S. Z. Gurbuz, H. D. Griffiths, A. Charlish, M. Rangaswamy, M. S. Greco and K. Bell, "An Overview of Cognitive Radar: Past, Present, and Future," in IEEE Aerospace and Electronic Systems Magazine, vol. 34, no. 12, pp. 6-18, 1 Dec. 2019.
\bibitem{Hay} S. Haykin, Y. Xue and P. Setoodeh, "Cognitive Radar: Step Toward Bridging the Gap Between Neuroscience and Engineering," in Proceedings of the IEEE, vol. 100, no. 11, pp. 3102-3130, Nov. 2012.
\bibitem{Haim} A. M. Haimovich, R. S. Blum and L. J. Cimini, "MIMO Radar with Widely Separated Antennas," in IEEE Signal Processing Magazine, vol. 25, no. 1, pp. 116-129, 2008.
\bibitem{li} J. Li and P. Stoica, "MIMO Radar with Colocated Antennas," in IEEE Signal Processing Magazine, vol. 24, no. 5, pp. 106-114, Sept. 2007.
\bibitem{aya1} A. M. Ahmed, S. Fortunati, A. Sezgin, M. S. Greco and F. Gini, "Robust Reinforcement Learning-based Wald-type Detector for Massive MIMO Radar," 2021 29th European Signal Processing Conference (EUSIPCO), Dublin, Ireland, 2021, pp. 846-850.

\bibitem{aya2} A. M. Ahmed, A. A. Ahmad, S. Fortunati, A. Sezgin, M. S. Greco and F. Gini, "A Reinforcement Learning Based Approach for Multitarget Detection in Massive MIMO Radar," in IEEE Transactions on Aerospace and Electronic Systems, vol. 57, no. 5, pp. 2622-2636, Oct. 2021.

\bibitem{fort1} S. Fortunati, L. Sanguinetti, F. Gini, M. S. Greco, and B. Himed , "Massive MIMO Radar for Target Detection"
IEEE Trans. Signal Process., vol. 68, pp. 859–871, Jan. 2020.

\bibitem{fried} B. Friedlander, "On Transmit Beamforming for MIMO Radar," in IEEE Transactions on Aerospace and Electronic Systems, vol. 48, no. 4, pp. 3376-3388, October 2012.
\bibitem{Tan} W. Tan, S. D. Assimonis, M. Matthaiou, Y. Han, X. Li and S. Jin, "Analysis of Different Planar Antenna Arrays for mmWave Massive MIMO Systems," 2017 IEEE 85th Vehicular Technology Conference (VTC Spring), Sydney, NSW, Australia, 2017.
\bibitem{Poole} D. Poole and A. Mackworth, Artificial Intelligence: Foundations of Computational Agents,
2nd ed. Cambridge, U.K.: Cambridge Univ. Press, 2017, [Online]. Available: http://artint.info/2e/html/ArtInt2e.html.

\bibitem{stoica} P. Stoica, J. Li and Y. Xie, "On Probing Signal Design For MIMO Radar," in IEEE Transactions on Signal Processing, vol. 55, no. 8, pp. 4151-4161, Aug. 2007.
\bibitem{fort2} S. Fortunati, F. Gini, M. S. Greco and C. D. Richmond, "Performance Bounds for Parameter Estimation under Misspecified Models: Fundamental Findings and Applications," in IEEE Signal Processing Magazine, vol. 34, no. 6, pp. 142-157, Nov. 2017.
\bibitem{sutton} R. S. Sutton and A. G. Barto, "Reinforcement Learning: An Introduction", 2nd ed. Cambridge, MA, USA: MIT Press, 2018, [Online], Available: http:// incompleteideas.net/book/the-book-2nd.html.
\bibitem{Swerling} P. Swerling, "Probability of Detection for Fluctuating Targets," in IRE Transactions on Information Theory, vol. 6, no. 2, pp. 269-308, April 1960.
\bibitem{nuttall} A. H. Nuttall, “Some Integrals Involving the (q sub m)-Function,” 1974.
\bibitem{dudgeon} Dan E. Dudgeon and Russell M. Mersereau, "Multidimensional Digital Signal Processing", Prentice Hall Professional Technical Reference, 1990.
\end{thebibliography}
\end{document}